\documentclass[12pt,oneside, a4paper]{article}

\ifx\pdfoutput\undefined
\usepackage[dvips,bookmarks=false]{hyperref}	
\else
\usepackage{hyperref}	
\fi
\hypersetup{colorlinks,bookmarksopen,bookmarksnumbered,citecolor=blue,
linkcolor=black,pdfstartview=FitH,urlcolor=blue}

\usepackage{graphicx}
\usepackage{amssymb}
\usepackage{cite}
\usepackage{bm}
\usepackage{indentfirst}
\usepackage{amsmath}
\usepackage{hhline}
\usepackage{multirow}

\allowdisplaybreaks

\newcommand{\Slash}[1]{{\ooalign{\hfil/\hfil\crcr$#1$}}}

\renewcommand{\bar}{\overline}

\usepackage{ulem}

\begin{document}

\begin{titlepage}

\begin{flushright}
 KUNS-2782\\
 KYUSHU-HET-203
\end{flushright}

\begin{center}

\vspace{1cm}
{\large\textbf{
Pseudo-Nambu-Goldstone dark matter\\ from gauged $U(1)_{B-L}$ symmetry
}
 }
\vspace{1cm}

\renewcommand{\thefootnote}{\fnsymbol{footnote}}
Yoshihiko Abe$^{1}$\footnote[1]{y.abe@gauge.scphys.kyoto-u.ac.jp}
,
Takashi Toma$^{2,3}$\footnote[2]{toma@staff.kanazawa-u.ac.jp}
,
Koji Tsumura$^{4}$\footnote[3]{tsumura.koji@phys.kyushu-u.ac.jp}
\vspace{5mm}

\textit{
$^1${Department of Physics, Kyoto University, Kyoto 606-8502, Japan}\\
 $^2${Department of Physics, McGill University,\\
 3600 Rue University, Montr\'{e}al, Qu\'{e}bec H3A 2T8, Canada}\\
 $^3${Institute of Liberal Arts and Science,\\
 Kanazawa University, Kakuma-machi, Kanazawa, 920-1192 Japan}\\
 $^4${Department of Physics, Kyushu University,\\
 744 Motooka, Nishi-ku, Fukuoka, 819-0395, Japan}
}

\vspace{8mm}

\abstract{
 A pseudo-Nambu-Goldstone boson (pNGB) is an attractive candidate for 
 dark matter since the current severe limits of dark matter direct detection 
 experiments are naturally evaded by its nature. 
 We construct a model with pNGB dark matter based on 
 a {\it gauged} $U(1)_{B-L}$ symmetry, 
 where no ad-hoc global symmetry is assumed. 
 The model keeps natural suppression mechanism for 
 the dark matter direct detection cross section. 
 On the other hand, 
 the pNGB can decay through the new high scale suppressed operators.  
 We show that the pNGB has long enough lifetime to be a dark matter 
 in the wide range of the parameter space of the model. 
The thermal relic abundance of pNGB dark matter can be fit with the observed value 
against the constraints on the dark matter decays from the cosmic-ray observations. 
}

\end{center}
\end{titlepage}

\renewcommand{\thefootnote}{\arabic{footnote}}
\newcommand{\bhline}[1]{\noalign{\hrule height #1}}
\newcommand{\bvline}[1]{\vrule width #1}

\setcounter{footnote}{0}

\setcounter{page}{1}

\section{Introduction}
\label{sec:1}
The existence of dark matter is inferred from 
various observations through gravity over the past
decades such as rotational curves of spiral
galaxies~\cite{Corbelli:1999af, Sofue:2000jx}, gravitational 
lensing~\cite{Massey:2010hh}, cosmic microwave
background~\cite{Aghanim:2018eyx} and collision of Bullet
Cluster~\cite{Randall:2007ph}.
However, the nature of dark matter is still unknown.
Identification of dark matter is important not only for cosmology but
also for particle physics because any standard model particles cannot 
play a role of  dark matter. 
Many kinds of dark matter candidates have been proposed so far.  
One of the prominent candidates is so-called Weakly Interacting Massive Particle (WIMP). 
The attractive feature of WIMPs is that the relic abundance is thermally
determined in the early universe.
The WIMP mass whose interaction is close the electroweak interaction is predicted in the range of $10$ GeV -- $100$ TeV.
Such WIMPs are basically detectable through non-gravitational interactions.
Although WIMPs are being searched through direct detection, indirect
detection and collider production, no clear signals of WIMPs have 
been confirmed yet. 
As a result, these experiments severely constrain WIMP mass and
interactions. 
In particular, recent direct detection experiments provide the strong
upper bounds on the elastic scattering cross section between dark matter
and nucleon~\cite{Akerib:2017kat,Cui:2017nnn,Aprile:2018dbl}.
In order to pursue WIMPs further in the current situation, we have to consider
mechanisms to avoid the severe constraint from the direct detection experiments. 
One way is to consider a fermionic dark matter with pseudo-scalar interactions \cite{Freytsis:2010ne}. 
In this case, since the scattering amplitude at tree level is suppressed by
the momentum transfer in non-relativistic limit due to the spin
structure, the leading contribution to the amplitude appears at loop
level~\cite{Ipek:2014gua,Arcadi:2017wqi,Sanderson:2018lmj,Abe:2018emu,Abe:2019wjw}.
Another option is to consider a pseudo-Nambu-Goldstone boson (pNGB) as
dark matter~\cite{Barger:2010yn,Gross:2017dan}.
Since all the interactions are written by derivative couplings in
non-linear representation, the scattering amplitude for direct detection
vanishes in non-relativistic limit.\footnote{A pNGB
dark matter also appears in the composite Higgs models. In this context,
the suppression of the elastic scattering amplitude has been studied in
Ref.~\cite{Fonseca:2015gva, Brivio:2015kia, Barducci:2016fue,
Balkin:2017aep, Balkin:2018tma, Ruhdorfer:2019utl, Ramos:2019qqa}.} 
The leading contribution comes from one-loop level, and the order of the
elastic cross section has been evaluated as $\mathcal{O}(10^{-48}) \
\mathrm{cm}^2$ at most~\cite{Azevedo:2018exj,Ishiwata:2018sdi}. 
Since this magnitude of the elastic cross section is considerably small,
probing pNGB dark matter by future direct detection experiments may be difficult. 
However, indirect detection and collider searches are more
promising, and there are some works in this direction~\cite{Huitu:2018gbc,Cline:2019okt}.
In addition, global fitting of the pNGB dark matter with comprehensive
analysis has been done in Ref.~\cite{Arina:2019tib}. 
In this paper, 
we propose a model of the pNGB dark matter from a {\it gauged} $U(1)_{B-L}$ symmetry.\footnote{
Gauge symmetries are also motivated by the conjecture that there is no global symmetry in quantum gravity~\cite{Banks:1988yz,Banks:2010zn}.}
We introduce two complex scalars with $Q_{B-L}=+1$ and $+2$, 
and three right-handed neutrinos for gauge anomaly cancellation. 
The pNGB dark matter scenario in Ref.~\cite{Gross:2017dan} is realized 
in the decoupling limit, where the $U(1)_{B-L}$ symmetry breaking scale 
is taken to be infinity. 
In contrast to the original pNGB dark matter scenario, 
the pNGB decays due to the new interactions through the heavy particles. 
The stability of the pNGB is determined by the breaking scale of 
the $U(1)_{B-L}$ symmetry. 
We show that the pNGB can be long-lived over the current upper bound 
of the lifetime from the cosmic-ray observations. 
We also study the consistencies with the relic abundance of dark matter, 
and low energy phenomenology. 

The rest of this paper is organized as follows. 
In Sec.~2, a pNGB is introduced from the $U(1)_{B-L}$ symmetry breaking. 
In Sec.~3, the longevity of the pNGB as dark matter is investigated. 
We also study the relevant constraints on our pNGB dark matter 
such as the relic abundance of dark matter, the perturbative unitarity, 
and the Higgs invisible decay and signal strength. 
Sec.~4 is devoted to our conclusion.

\section{The Model}
\label{sec:2}
\begin{table}[t]
\centering
\begin{tabular}{|c||c|c|c|c|c|c||c|c|c|}\hline
 & $Q_L$ & $u_{R}^{c}$ & $d_{R}^{c}$ & $L$ & $e_{R}^{c}$ & $H$ & $\nu_{R}^c$ & $S$ & $\Phi$ \\
\hhline{|=#=|=|=|=|=|=#=|=|=|}
 $SU(3)_{c}$ & $\bm{3}$ & $\bar{\bm{3}}$ & $\bar{\bm{3}}$ & $\bm{1}$ & $\bm{1}$ & $\bm{1}$ &
			 $\bm{1}$ & $\bm{1}$ & $\bm{1}$ \\ \hline
 $SU(2)_{L}$ & $\bm{2}$ & $\bm{1}$ & $\bm{1}$ & $\bm{2}$ & $\bm{1}$ & $\bm{2}$ &
			 $\bm{1}$ & $\bm{1}$ & $\bm{1}$ \\ \hline
 $U(1)_{Y}$ & $+1/6$ & $-2/3$ & $+1/3$ & $-1/2$ & $+1$ & $+1/2$ &
			 $0$ & $0$ & $0$ \\\hline
 $U(1)_{B-L}$ & $+1/3$ & $-1/3$ & $-1/3$ & $-1$ & $+1$ & $0$ &
			 $+1$ & $+1$ & $+2$ \\\hline
\end{tabular}
\caption{Particle contents and quantum charges.}
\label{tab:1}
\end{table}

The particle contents and the charge assignments under the gauge group $SU(3)_{c} \times SU(2)_{L} \times U(1)_{Y} \times U(1)_{B-L}$ are shown in Tab.~\ref{tab:1}.
We note that the model is consist of particles in the ordinary $U(1)_{B-L}$ model
and an additional scalar singlet $S$ with $Q_{B-L}=+1$. 
The gauge kinetic terms of the new particles charged under $U(1)_{B-L}$ are written as
\begin{align}
 \mathcal{L}_{K}= (D_\mu S)^\dagger (D^\mu S) + (D_\mu \Phi)^\dagger (D^\mu \Phi) + \bar{\nu_{R}} i \Slash{D} \nu_{R} -\frac{1}{4}X_{\mu\nu}X^{\mu\nu} -\frac{\sin \epsilon}{2}X_{\mu\nu}B^{\mu\nu},
\end{align}
where $D_\mu \equiv \partial_\mu +i g_{B-L} Q_{B-L}X_\mu$ is the
covariant derivative with the new gauge boson $X_\mu$ associated with the
$U(1)_{B-L}$ symmetry. 
The field strengths for $U(1)_{B-L}$ and $U(1)_{Y}$ are denoted by
$X_{\mu\nu}$ and $B_{\mu\nu}$, respectively. 
The last term is the gauge kinetic mixing between $X_\mu$ and $B_\mu$. 
An extra mass eigenstate $Z^\prime$ of neutral gauge bosons 
is mainly composed by the new gauge boson $X_{\mu}$. 
The detailed calculations of diagonalization of the kinetic mixing and
mass matrix is summarized in Appendix~\ref{sec:Gauge kinetic mixing}. 
\medskip

The scalar potential 
is written as
%
\begin{align}
 V(H,S,\Phi) =& -\frac{\mu_{H}^2}{2}|H|^2 -\frac{\mu_{S}^2}{2} |S|^2
 -\frac{\mu_{\Phi}^2}{2}|\Phi|^2 +\frac{\lambda_{H}}{2} |H|^4
 +\frac{\lambda_{S}}{2}|S|^4 + \frac{\lambda_{\Phi}}{2}|\Phi|^4 
 \nonumber\\
 &+\lambda_{HS} |H|^2 |S|^2 + \lambda_{H\Phi}|H|^2|\Phi|^2 +\lambda_{S\Phi}|S|^2|\Phi|^2
 -\biggl( \frac{\mu_{\mathrm{c}}}{\sqrt{2}} \Phi^* S^2 + \mathrm{c.c.}\biggr).
 \label{eq:potential}
\end{align}
%
The CP phase of the 
cubic term is eliminated by the field redefinition of $\Phi$.
All the scalar fields develop vacuum expectation values (VEVs), and 
they are parametrized by
%
\begin{align}
 H= \left( \begin{array}{c}
 0 \\
 (v+h)/\sqrt{2}
 \end{array}\right),
 \hspace{1em}
 S=\frac{v_{s} + s+ i \eta_{s}}{\sqrt{2}},
 \hspace{1em}
 \Phi= \frac{v_\phi + \phi + i\eta_{\phi}}{\sqrt{2}}.
\end{align}
%
In the limit $\mu_\mathrm{c}\to0$, 
the scalar potential has two independent global $U(1)$ symmetries associated
with the phase rotation of $S$ and $\Phi$, respectively.
When $\mu_{\mathrm{c}} \ne 0$, these $U(1)$ symmetries are 
merged to the $U(1)_{B-L}$ symmetry. 
Therefore, one of NGBs is absorbed by $X_\mu$, while the other appears 
as a physical pNGB with the mass proportional to $\mu_{\mathrm{c}}$. 
We note that $\mu_\mathrm{c}$ is naturally small in 't Hooft sense 
because of the enhanced symmetry argument. 
One can intuitively understand that if the scalar $\Phi$ gets the VEV $v_{\phi}$ 
, the last term 
gives effective mass term 
$\mu_{\mathrm{c}} v_{\phi} S^2 /2
$ for the NGB. 
By solving stationary conditions for $\mu_{H}^2, \mu_{S}^2, \mu_{\Phi}^2$, 
the mass matrix for the CP-even scalars in the
$(h, s, \phi)$ basis is
%
\begin{align}
 M_{\mathrm{even}}^2 =&
 \left( \begin{array}{ccc}
 \lambda_H v^2 & \lambda_{HS} v v_{s} & \lambda_{H\Phi}v v_{\phi} \\
 \lambda_{HS} v v_{s} & \lambda_S v_{s}^2 & \lambda_{S\Phi}v_{s} v_{\phi}-\mu_{\mathrm{c}} v_{s} \\
 \lambda_{H\Phi} v v_{\phi} & \lambda_{S\Phi} v_{s} v_{\phi} -\mu_{\mathrm{c}} v_{s} & \lambda_{\Phi} v_{\phi}^2 +\frac{\mu_{\mathrm{c}} v_{s}^2}{2v_{\phi}}
 \end{array}\right). 
\end{align}
%
This mass matrix is approximately diagonalized by the matrix
%
\begin{align}
 U \approx 
 \left( \begin{array}{ccc}
 1 & 0 & \frac{\lambda_{H\Phi}v}{\lambda_{\Phi} v_{\phi}} \\
 0 & 1 & \frac{\lambda_{S\Phi}v_{s}}{\lambda_{\Phi} v_{\phi}} \\
 -\frac{\lambda_{H\Phi}v}{\lambda_{\Phi} v_{\phi}} & -\frac{\lambda_{S\Phi}v_s}{\lambda_{\Phi}v_{\phi}}& 1
 \end{array}\right)
 \left( \begin{array}{ccc}
 \cos \theta & \sin \theta & 0 \\
 - \sin \theta & \cos \theta & 0 \\
 0 & 0 & 1
 \end{array}\right),
 \label{eq:Umix}
\end{align}
%
where $U(1)_{B-L}$ symmetry breaking is assumed mainly by $v_\phi$. 
The gauge eigenstates $(h, s, \phi)$ are expressed by 
the mass eigenstates $(h_1, h_2, h_3)$ as
%
\begin{align}
 \left( \begin{array}{c}
 h \\
 s \\
 \phi 
 \end{array}\right)
 = U
 \left( \begin{array}{c}
 h_1 \\
 h_2 \\
 h_3
 \end{array}\right),
\end{align}
%
where the mixing angle $\theta$ is given by
%
\begin{align}
 \tan 2 \theta ~ \approx ~\frac{2 v v_{s} (\lambda_{HS} \lambda_{\Phi} -\lambda_{H\Phi} \lambda_{S\Phi})}{ v^2 ( \lambda_{H\Phi}^2 -\lambda_{H} \lambda_{\Phi}) -v_{s}^2 (\lambda_{S\Phi}^2 -\lambda_{S} \lambda_{\Phi})}.
\end{align}
%
The corresponding mass eigenvalues for $h_i$ are approximately evaluated
as
%
\begin{align}
 m_{h_1}^2 \approx &~ \lambda_{H} v^2 - \frac{ \lambda_{H\Phi}^2 \lambda_{S} -2 \lambda_{HS} \lambda_{H\Phi} \lambda_{S\Phi} + \lambda_{\Phi} \lambda_{HS}^2}{\lambda_{S} \lambda_{\Phi} -\lambda_{S\Phi}^2 } v^2,
 \label{eq:massh1}
 \\
 m_{h_2}^2 \approx &~ \frac{\lambda_{S}\lambda_{\Phi} -\lambda_{S\Phi}^2}{\lambda_{\Phi}} v_{s}^2 +\frac{(\lambda_{\Phi} \lambda_{HS}-\lambda_{H\Phi} \lambda_{S\Phi})^2}{\lambda_{\Phi}(\lambda_{S} \lambda_{\Phi} -\lambda_{S\Phi}^2)} v^2,
 \label{eq:massh2}
 \\
 m_{h_3}^2 \approx &~ \lambda_{\Phi} v_{\phi}^2.
\end{align}
%
We identify $h_1$ as the SM-like Higgs boson with the mass $m_{h_1}=125\ \mathrm{GeV}$.
The mass matrix of the CP-odd scalars in the gauge eigenstates $(\eta_{s}, \eta_{\phi})$ is written as
%
\begin{align}
 M_{\mathrm{odd}}^2 = \frac{\mu_{\mathrm{c}}}{2v_{\phi}} 
 \left( \begin{array}{cc}
 4v_{\phi}^2 & -2 v_{s} v_{\phi} \\
 -2 v_{s} v_{\phi} & v_{s}^2
 \end{array}\right).
\end{align}
%
This mass matrix can be diagonalized as
%
\begin{align}
 V^{\mathrm{T}} M_{\mathrm{odd}}^2 V = \left( \begin{array}{cc}
 m_{\chi}^2 & 0 \\
 0 & 0
 \end{array}\right),
 \hspace{1em}
 m_{\chi}^2 = \frac{(v_{s}^2 + 4 v_{\phi}^2 ) \mu_{\mathrm{c}} }{4 v_{\phi}},
\end{align}
%
where the unitary matrix $V$ is given by
%
\begin{align}
 V= \frac{ 1 }{ \sqrt{ v_{s}^2 + 4 v_{\phi}^2 }} \left( \begin{array}{cc}
 2 v_{\phi} & v_s \\
 -v_{s} & 2 v_{\phi}
 \end{array}\right).
\end{align}
%
The gauge eigenstates $( \eta_{s} , \eta_{\phi} )$ are rewritten 
by the mass eigenstates $( \chi , \tilde{\chi} )$ as
%
\begin{align}
 \left( \begin{array}{c}
 \eta_{s} \\
 \eta_{\phi} \\
 \end{array}\right)
 = \frac{1}{\sqrt{ v_{s}^2 + 4 v_{\phi}^2 }} \left( \begin{array}{cc}
 2 v_{\phi} & v_{s} \\
 -v_{s} & 2 v_{\phi}
 \end{array}\right)
 \left( \begin{array}{c}
 \chi \\
 \tilde{\chi} 
 \end{array}\right),
 \label{eq:Vmix}
\end{align}
%
where $\tilde{\chi}$ is the NGB absorbed by $X_\mu$, 
and $\chi$ corresponds to the pNGB 
which will be identified as dark matter. 

\medskip
The following Yukawa interactions are also invariant under the
imposed symmetry
%
\begin{align}
 \mathcal{L}_{Y}= - (y_{\nu})_{ij} \tilde{H}^\dagger \bar{\nu_{Ri}} L_j -\frac{(y_{\Phi})_{ij}}{2} \Phi \bar{\nu_{Ri}^c} \nu_{Rj} + \mathrm{h.c.}
\end{align}
%
After the $U(1)_{B-L}$ symmetry breaking, 
the right-handed neutrinos obtain the Majorana mass 
$M\equiv y_{\Phi} v_{\phi} /\sqrt{2}$.
Thus,  the small masses for active neutrinos are generated 
by the type-I seesaw mechanism as 
$m_{\nu} \approx - m_{D} M^{-1} m_{D}^{\mathrm{T}}$ with
the Dirac mass 
$m_{D} \equiv y_{\nu} v /\sqrt{2}$. 
Since the heaviest neutrino mass is roughly fixed by the neutrino
oscillation data as $m_{\nu} \sim
0.1\ \mathrm{eV}$, the required scale of the VEV $v_{\phi}$ is estimated
as 
%
\begin{align}
 m_{\nu} \sim \frac{y_{\nu}^2 v^2}{\sqrt{2} y_{\Phi} v_{\phi}} \sim 0.1 \ \mathrm{GeV}
 \hspace{1em}
 \to
 \hspace{1em}
 v_{\phi} \sim 4.3 \times 10^{14}\, \mathrm{GeV} \biggl( \frac{y_{\nu}^2}{y_{\Phi}}\biggr).
\end{align}
%
The scale 
$v_\phi$ is large enough as compared to the electroweak scale 
unless the Dirac Yukawa coupling $y_{\nu}$ is considerably small. 
%

\section{Long-lived Dark Matter}

%
First of all, we check the cancellation of the scattering amplitude for
direct detection in this model.
When $v_\phi$ is much larger than $v$ and $v_s$, 
the three-point interactions among the pNGB and CP-even scalars are expressed as 
%
\begin{align}
 \mathcal{L}_{\chi\chi h_i} = -\sum_{i =1,2,3} \frac{\kappa_{\chi\chi h_i}}{2} \chi^2 h_i,
\end{align}
%
where each coupling coefficient $\kappa_{\chi \chi h_i}$ is given by
%
\begin{align}
 &\kappa_{\chi \chi h_1} \approx  -\frac{m_{h_1}^2\sin \theta}{v_s},
 \hspace{1em}
 \kappa_{\chi \chi h_2} \approx + \frac{m_{h_2}^2 \cos \theta}{v_s},
 \hspace{1em}
 \kappa_{\chi \chi h_3} \approx + \frac{m_{h_3}^2}{v_s} \frac{\lambda_{S\Phi}v_s}{\lambda_\Phi v_\phi}.
\end{align}
%
We note that these couplings are proportional to the corresponding scalar masses.
The CP-even scalar exchanging scattering amplitudes of the pNGB and SM particles are expressed as
%
\begin{align} 
 i \mathcal{M} \propto 
 \frac{\sin \theta \cos \theta}{v_s} \biggl( - \frac{m_{h_1}^2}{q^2 -m_{h_1}^2} + \frac{m_{h_2}^2}{q^2 -m_{h_2}^2} \biggr)
 +{\mathcal O}(1/v_\phi), 
\end{align}
%
where $q$ is the momentum transfer.
Due to this structure, the elastic scattering cross section 
of dark matter and nucleon is suppressed in the non-relativistic limit.
This is nothing less than the same cancellation mechanism of the pNGB dark matter 
for the direct detection\cite{Gross:2017dan}.\footnote{
This cancellation mechanism works if and only if the $U(1)_{B-L}$ charge of $S$ 
is unity.
}
Therefore, the pNGB derived from the gauged $U(1)_{B-L}$ model 
can be a good candidate for dark matter. 
\medskip

It is necessary to examine the longevity of the pNGB to be dark matter,  
because our pNGB is unstable. 
The SM particles are produced by the decays of the pNGB dark matter candidate, 
 and these particles further decay into the stable particles such
as $e^{\pm}$, $\gamma$, $\nu$, $p$, $\bar{p}$. 
These cosmic-rays can be signals of dark matter or constrained by observations. 
In this paper, following the analysis of gamma rays coming from dwarf
spheroidal galaxies using Fermi-LAT data~\cite{Baring:2015sza}, 
we study constraints of our model from a conservative limit of the dark matter lifetime 
$\tau_{\mathrm{DM}} \gtrsim 10^{27}\ \mathrm{s}$, or equivalently
$\Gamma_{\mathrm{DM}} \lesssim 6.6 \times 10^{-52}\ \mathrm{GeV}$ in
terms of decay width.
\medskip

\begin{figure}[t]
\centering
 \includegraphics[scale=0.9]{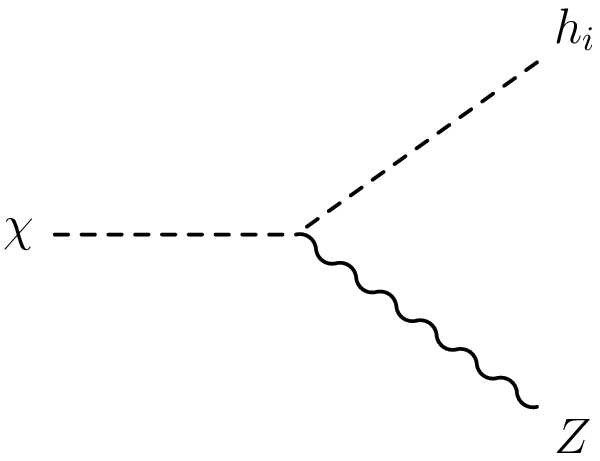}
 \hspace{1cm}
 \includegraphics[scale=0.9]{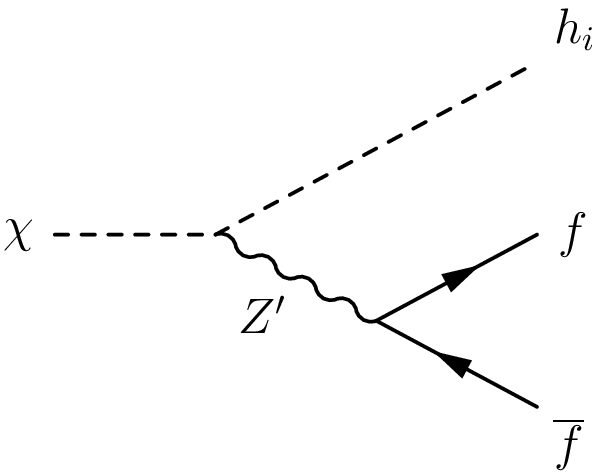}
 \caption{Feynman diagrams of the dark matter decay.}
 \label{fig:decay}
\end{figure}

One of possible two body decay channels is $\chi \to \nu \nu$ through
the scalar mixing and the neutrino heavy-light mixing. 
The partial decay width is roughly estimated as
$\Gamma_{\chi \to \nu \nu} \lesssim 10^{-67} \ \mathrm{GeV}$. 
This is small enough to guarantee the dark matter (meta-)stability 
thanks to the strong suppression by the small neutrino masses.
In addition, 
the current experimental upper bound for this channel is much weaker than 
our estimate, 
since the observation of the produced neutrino cosmic-rays is much
more difficult than those of charged particles such as $e^{\pm}$, $p$, 
$\bar{p}$. 
Thus, this decay channel can be safely ignored.
Another two body decay mode $\chi \to h_i Z$, depicted in the left panel of
Fig.~\ref{fig:decay}, becomes important if it is kinematically allowed for  
$m_{\chi} > m_{h_i}+m_Z$.
The total decay width for this channel is computed as
%
\begin{align}
 \Gamma_{\text{2-body}}=&~ \sum_{i}  \Gamma_{\chi \to h_i Z} \approx \frac{g_{B-L}^2}{16 \pi m_{Z'}^4 } m_{Z}^2 m_{\chi}^3 \sin^2 \theta_{W} \sin^2 \epsilon
 \nonumber\\
 =&~ 
 5.8 \times 10^{-52}\ \mathrm{GeV} \biggl( \frac{m_{\chi}}{0.5 \  \mathrm{TeV}} \biggr)^3 \biggl( \frac{10^{15}\ \mathrm{GeV}}{m_{Z'}} \biggr)^2 \biggl( \frac{10^{15}\ \mathrm{GeV}}{v_\phi} \biggr)^2 \biggl( \frac{\sin \epsilon}{1/\sqrt{2}} \biggr)^2,
 \label{eq:two}
\end{align}
%
where the mass hierarchy $m_{h_1},m_{h_2}, m_{Z} \ll m_{\chi}\ll
m_{h_3},m_{Z^\prime}$ is applied for this approximated formula. 
This two body decay becomes important if there is a large gauge kinetic mixing $\sin \epsilon$,\footnote{
A bound on the kinetic mixing $\sin\epsilon$ 
is obtained from the perturbative unitarity if the new gauge
boson mass is lighter than TeV scale~\cite{Bandyopadhyay:2018cwu}.
However this bound is irrelevant to our case since the new
gauge mass is assumed to be much heavier than TeV scale.
} and is irrelevant for the vanishing gauge-kinetic mixing.
Since $\chi$ is $\eta_s$-like, the main contribution to this decay channel 
comes from $\chi \to h_2 Z$ where $h_2$ is $s$-like. 
Further suppression due to the scalar mixing is expected for other decay channels, 
e.g., $\chi \to h_1 Z$. 
A decay process emitting a photon such as $\chi \to h_{i} \gamma$ is forbidden due to the helicity conservation.

One can naively expect that three body decay processes are subdominant if
the above two body decay processes are kinematically allowed.
However, three body decays could be dominant depending on parameters, in
particular when the gauge kinetic mixing is small. 
There are two possible three body decay processes $\chi \to Z f
\bar{f}$ and $\chi \to h_i f \bar{f}$.
The former is mediated by the heavy CP-even scalar $h_3$, and is
possible only when the gauge kinetic mixing is non-zero as same as the
above two body decay process. 
The decay width is extremely suppressed by the heavy $h_3$ mass and small
scalar mixing, thus this is ignored.
The latter process is mediated by the heavy $Z^\prime$ gauge boson as
depicted in the right panel of Fig.~\ref{fig:decay}. 
In the case that $m_{f} \ll m_{h_i}, m_{\chi} \ll m_{Z'}$, the decay
width is computed as
%
\begin{align}
 \Gamma_{\chi \to h_i f \bar{f}} = \frac{g_{B-L}^2 U_{si}^2 m_{\chi}^5}{768 \pi^3 m_{Z'}^4} \frac{\cos^2 \zeta}{\cos^2 \epsilon} \Bigl( {g^{f}_{V}}^2 + {g^{f}_{A}}^2 \Bigr) \Bigl[ 1 - 8\xi_i + 8 \xi_i^3 -\xi_i^4 -12 \xi_i^2 \log \xi_i\Bigr],
\end{align}
%
where $\xi_i \equiv m_{h_i}^2 / m_{\chi}^2$ and $U_{si}~(i=1,2)$ is the
element of the CP-even scalar mixing matrix in Eq.~(\ref{eq:Umix}). 
The mixing matrix elements are explicitly given by $U_{s1} \approx -\sin
\theta$, $U_{s2} \approx \cos \theta$.
The coefficients $g^{f}_{V/A}$ are the coupling constants between the heavy
gauge boson $Z'$ and 
vector or axial vector current, which are defined by
%
\begin{align}
 \mathcal{L}_{Z' \bar{f} f}= -Z'_\mu \bar{f} \gamma^\mu \Bigl[ g^f_V +g^f_A \gamma_5 \Bigr] f.
\end{align}
%
Their expressions in $m_{Z} \ll m_{Z'}$ limit are given by
%
\begin{align}
 g^f_V \approx &-g_1 (Q^f_{\mathrm{em}} -T^f_3 ) \tan \epsilon + \frac{g_{B-L}}{\cos \epsilon} Q^f_{B-L},
 \\
 g^f_A \approx & ~ 0,
\end{align}
%
%
%
where $Q^f_{\mathrm{em}}$, $T^f_{3}$ and $Q^f_{B-L}$ correspond to
electromagnetic charge, the third component of weak isospin and $B-L$ charge of the fermion $f$,
respectively.
The mixing angle $\zeta$ is introduced to diagonalize the gauge boson
mass matrix as summarized in Appendix~\ref{sec:Gauge kinetic mixing}.
It is useful to take some specific values of the parameters to
understand the behavior of the three body decay width.
Here, we consider the two cases, $\sin \epsilon=0$ and $1/\sqrt{2}$.
First, when there is vanishing gauge kinetic mixing ($\sin\epsilon=0$), 
the total three body decay width can simply be computed as
%
\begin{align}
 \Gamma_{\text{3-body}} \Bigr|_{\sin \epsilon \to 0 } = & \sum_{i} \sum_{f} \Gamma_{\chi \to h_i f \bar{f}} \approx \frac{13}{16} \frac{g_{B-L}^4}{1536 \pi^3} \frac{m_{\chi}^5}{m_{Z'}^4}
 \nonumber\\
 \approx & ~
5.3 \times 10^{-52} \ \mathrm{GeV} \biggl( \frac{m_{\chi}}{0.5\ \mathrm{TeV}} \biggr)^5 \biggl( \frac{10^{15} \ \mathrm{GeV}}{v_\phi} \biggr)^4,
 \label{eq:threeno}
\end{align}
%
where we used the relation $m_{Z'}^2 \approx 4 g_{B-L}^2 v_{\phi}^2$.
The second case is a typical value of non-zero gauge kinetic mixing ($\sin \epsilon = 1/\sqrt{2}$).
Then, the total decay width can be evaluated as
%
\begin{align}
 \Gamma_{\text{3-body}} \Bigl|_{\sin \epsilon \to 1/\sqrt{2}} =& \sum_{i} \sum_{f} \Gamma_{\chi \to h_i f \bar{f}} \approx \frac{g_{B-L}^2}{768 \pi^3} \frac{m_{\chi}^5}{m_{Z'}^4} \bigl( 10 g_{1}^{2} - 8\sqrt{2} g_{1} g_{B-L} + 26 g_{B-L}^2 \bigr)
 \nonumber\\
 =& ~
 4.1 \times 10^{-52} \ \mathrm{GeV} \biggl( \frac{m_{\chi}}{0.5\ \mathrm{TeV}} \biggr)^5 \biggl( \frac{10^{15} \ \mathrm{GeV}}{m_{Z'}} \biggr)^2 \biggl( \frac{10^{15}\ \mathrm{GeV}}{v_\phi} \biggr)^2
 \nonumber\\
 &\times 
 \biggl[ 1- \frac{2\sqrt{2}}{5} \frac{m_{Z'}}{g_1 v_\phi} + \frac{13}{20} \frac{m_{Z'}^2}{g_1^2 v_\phi^2}\biggr].
 \label{eq:threemax}
\end{align}
%
From the above calculations, one can find that the two body decay width
in Eq.~(\ref{eq:two}) is proportional to $m_{\chi}^3$, while the three
body decay widths in Eq.~(\ref{eq:threeno}), (\ref{eq:threemax}) are
proportional to $m_{\chi}^5$. 
Therefore the three body decay width tends to be dominant when the dark matter
mass $m_{\chi}$ is large.
Another important point is that the two body decay width vanishes when
there is no gauge kinetic mixing while the three body decay occurs even
in the case. 

\begin{figure}[t]
\centering
 \includegraphics[scale=0.66]{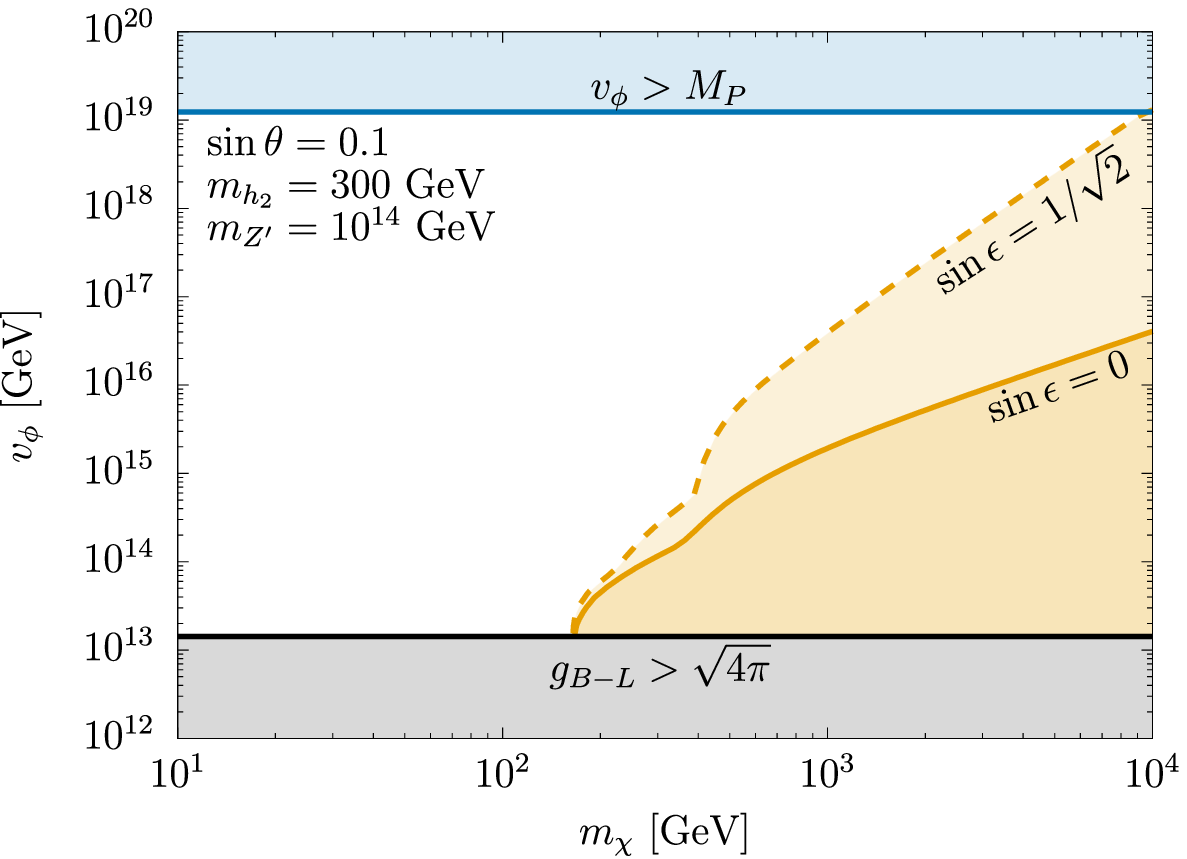}
 \includegraphics[scale=0.66]{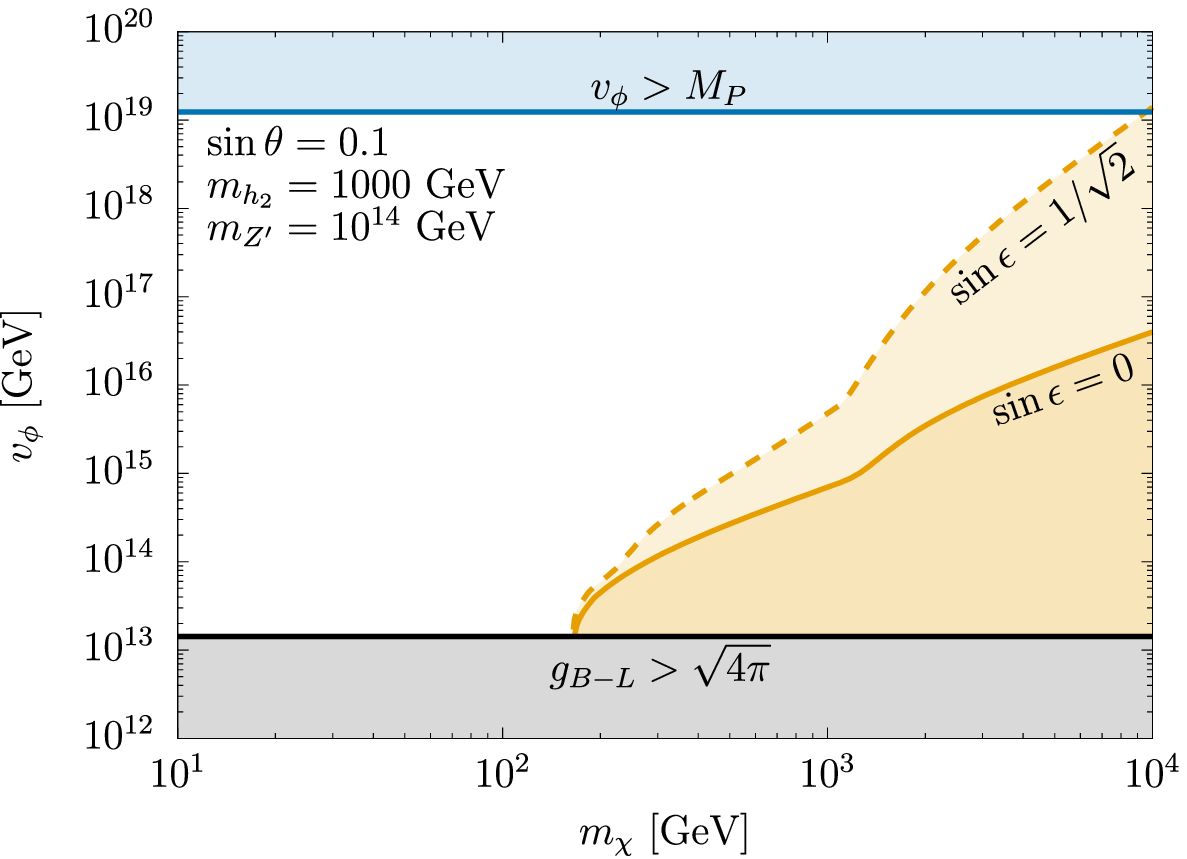}
 \\
 \includegraphics[scale=0.66]{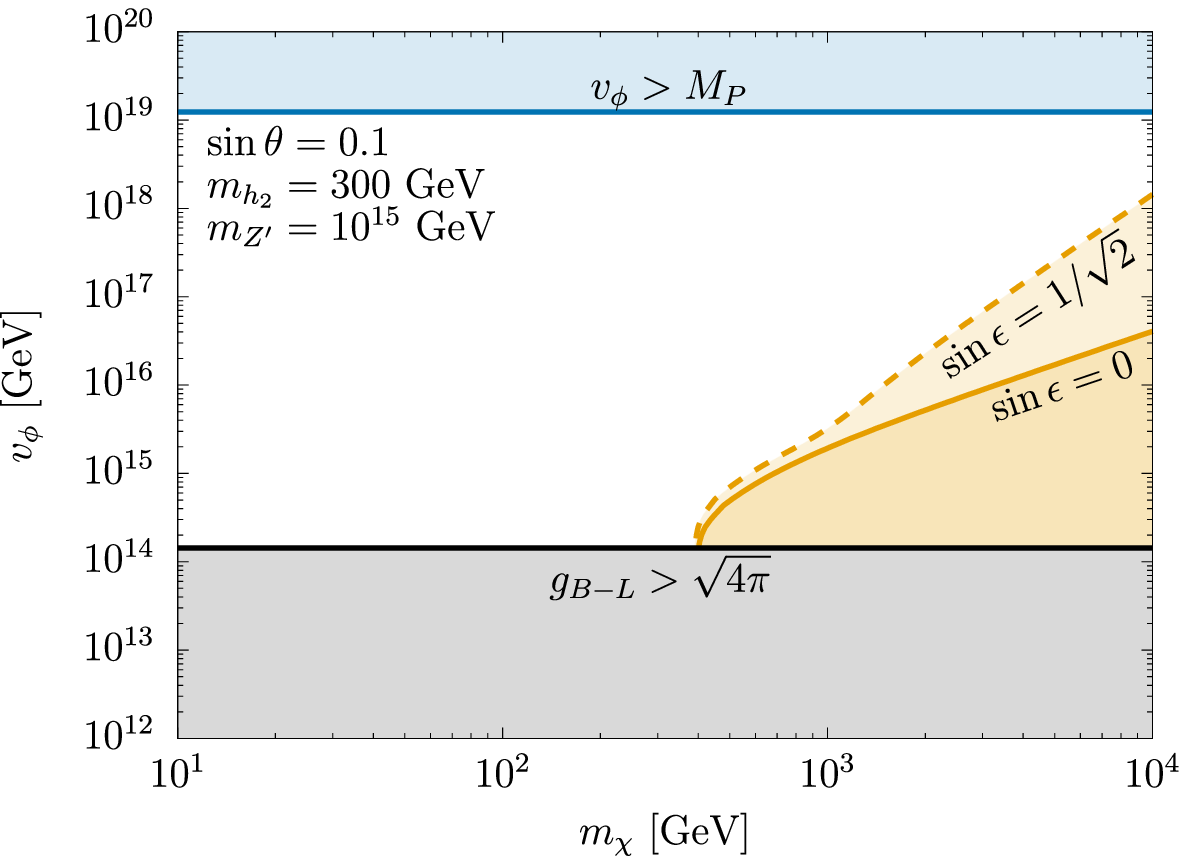}
 \includegraphics[scale=0.66]{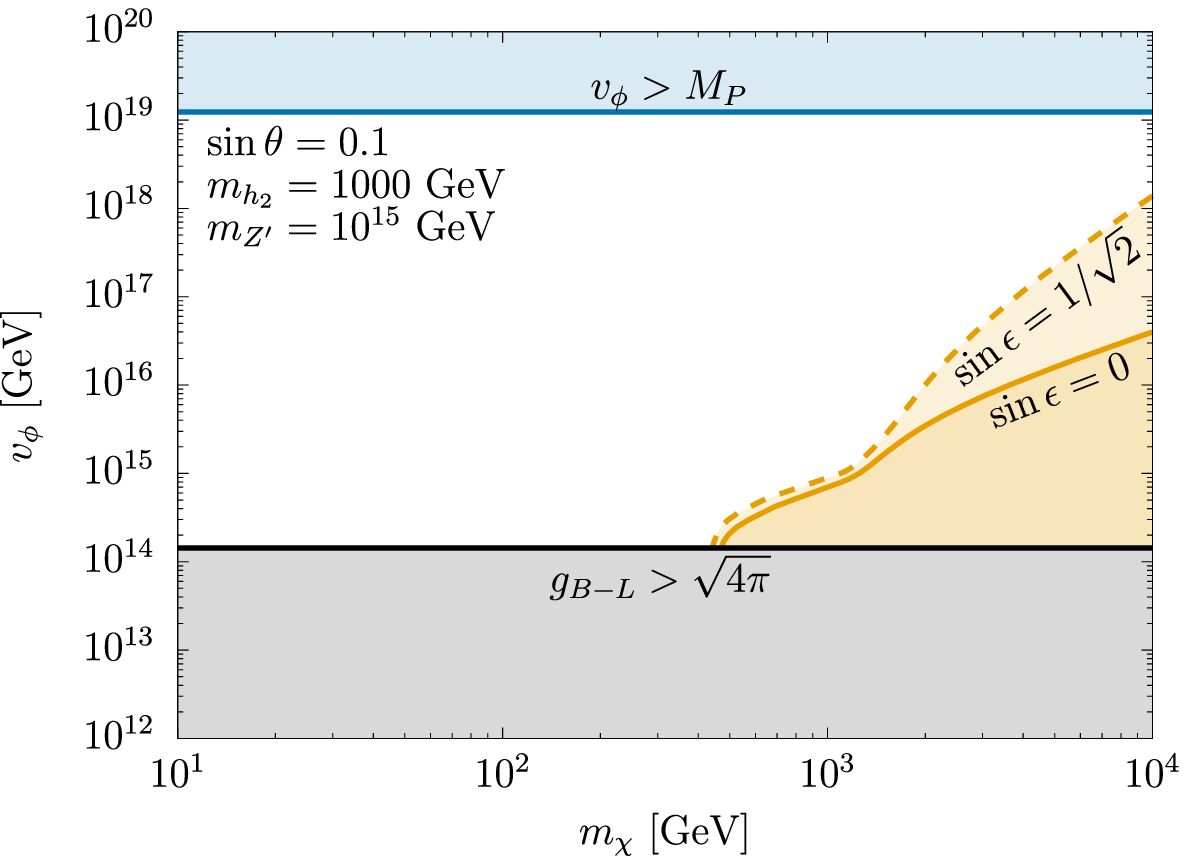}
\caption{
Allowed regions in the $(m_{\chi},  v_{\phi} )$ plane. 
The scalar mass is fixed as $m_{h_2} =300\ \mathrm{GeV}$ for the left panels and $m_{h_2}  = 1000 \ \mathrm{GeV}$ for the right panels.
The new gauge boson mass is fixed as $m_{Z'} = 10^{14}\ \mathrm{GeV}$ in the upper panels and $m_{Z'} = 10^{15} \ \mathrm{GeV}$ in the lower panels.
The orange regions are excluded by the conservative bound of
 the dark matter lifetime ($\tau_\mathrm{DM}\gtrsim10^{27}~\mathrm{s}$). 
The gray region is disfavored by the perturbative unitarity bound 
of the gauge coupling $g_{B-L}$. 
The upper light blue region denotes the parameter space that the VEV
 $v_\phi$ becomes larger than the Planck mass $M_P$. 
}
\label{fig:mx_vev_mzp}
\end{figure}
\medskip

In our model, there are $10$ independent parameters in total, which are
relevant to the decaying pNGB dark matter. 
These may be chosen to as
$m_{\chi},m_{h_2},m_{h_3},m_{Z^\prime},\sin\theta$, $v_s$, $v_\phi$, 
$\lambda_{H\Phi}$ ,$\lambda_{S\Phi}$ and $\sin\epsilon$.
The Yukawa couplings $y_{\nu}$ and $y_{\Phi}$ are irrelevant for the 
pNGB sector, and one can always take appropriate Yukawa couplings
and right-handed neutrino masses consistently with the neutrino oscillation data. 
Only $4$ parameters ($m_{\chi},\sin\theta,v_s,m_{h_2}$) are important
for the phenomena of the stable dark matter, 
which are used in the discussion in the next section. 
The other parameters are relevant to the dark matter decay.
In our numerical calculations, we choose the following parameter sets as examples:
%
 \begin{align}
  m_{h_2} = 300~\text{or}~1000\ \mathrm{GeV},\quad
  m_{h_3} = 10^{13}\ \mathrm{GeV},\quad
  m_{Z^\prime}=10^{14}~\text{or}~10^{15}~\mathrm{GeV},
  \nonumber\\
  \sin\theta=0.1,\quad
  \lambda_{H\Phi} =\lambda_{S\Phi} = 10^{-6},\quad
  \sin\epsilon=0~\text{or}~\frac{1}{\sqrt{2}}.
  \hspace{1.5cm}
  \label{eq:param_set}
 \end{align}
%
The gauge coupling $g_{B-L}$ and the quartic coupling $\lambda_\Phi$ are
fixed by $g_{B-L}\approx m_{Z^\prime}^2/(4v_\phi^2)$ and
$\lambda_\phi\approx m_{h_3}^2/v_\phi^2$ for a given VEV $v_\phi$. 
The mixing angle $\sin\theta$ is constrained as $\sin\theta\lesssim0.3$
for $m_{h_2}\gtrsim100~\mathrm{GeV}$ by the electroweak
precision measurements and the direct search of the second Higgs
boson~\cite{Martin-Lozano:2015dja, Falkowski:2015iwa}. 
This constraint can also be applied for our model. 
The quartic couplings $\lambda_{H\Phi}$ and $\lambda_{S\Phi}$ are taken
small such that the approximate formulae Eq.~(\ref{eq:massh1}) and
(\ref{eq:massh2}) are valid.
If these couplings are large, the negative contributions to the CP-even
scalar masses in Eq.~(\ref{eq:massh1}) and (\ref{eq:massh2}) become
significant and make them tachyonic. 
Note that one can take these quartic couplings larger than 
Eq.~(\ref{eq:param_set}) for smaller VEV $v_\phi$. 
However, we choose as Eq.~(\ref{eq:param_set}) for simplicity 
so that the quartic couplings retain constant in our numerical calculations.

In Fig.~\ref{fig:mx_vev_mzp}, 
we show the allowed parameter region from the (meta-)stability constraint 
of dark matter in the plane 
$(m_{\chi}, v_{\phi} )$. 
The orange region is ruled out  by the cosmic-ray observation. 
The perturbative unitarity bound of the $U(1)_{B-L}$ gauge coupling 
exclude the lower gray region. 
The VEV $v_{\phi}$ becomes larger
than the Planck scale $M_{P}=1.2\times10^{19}~\mathrm{GeV}$ in the upper
light blue region.\footnote{
If we consider a cosmic string creation after the inflation, 
the VEV breaking $U(1)_{B-L}$ symmetry is restricted as $v_{\phi} < 4 \times 10^{15}\  \mathrm{GeV}$ from the CMB observation, 
which is discussed in Ref.~\cite{Charnock:2016nzm}.
}
One can find from the plots that when the dark matter mass $m_{\chi}$ becomes larger
than the threshold of the decay channel $\chi \to h_2 f \bar{f}$
($m_{\chi}\gtrsim m_{h_2}$),
the total decay width is enhanced and the bound of the cosmic-ray
observations becomes stronger.
The scaling behavior of the orange region is observed as $v_{\phi}
\propto m_{\chi}^{5/4}$ for no kinetic mixing and $v_{\phi} \propto
m_{\chi}^{5/2}$ for a large kinetic mixing in heavier dark matter mass region.
This follows from the analytic formulae of the total three body decay
width in Eqs.~(\ref{eq:threeno}) and (\ref{eq:threemax}). 
Characteristic threshold behaviors are also seen at $m_\chi \sim m_{h_2}+m_Z$, 
where $m_{h_2}=300\ (1000)$ GeV is taken in the left (right) panels.  
We here comment on the possible four body decay channel. 
If the dark matter mass is too small to decay through the above two
or three body decay process, the four body
decay process $\chi \to h_{i}^{*} Z^{*} \to f \bar{f} 
f^\prime \bar{f^\prime}$ would be the main decay channel of dark matter. 
However, the decay width is too small to be constrained or be signals of
dark matter at present. 
\medskip

\begin{figure}[t]
\centering
 \includegraphics[scale=0.66]{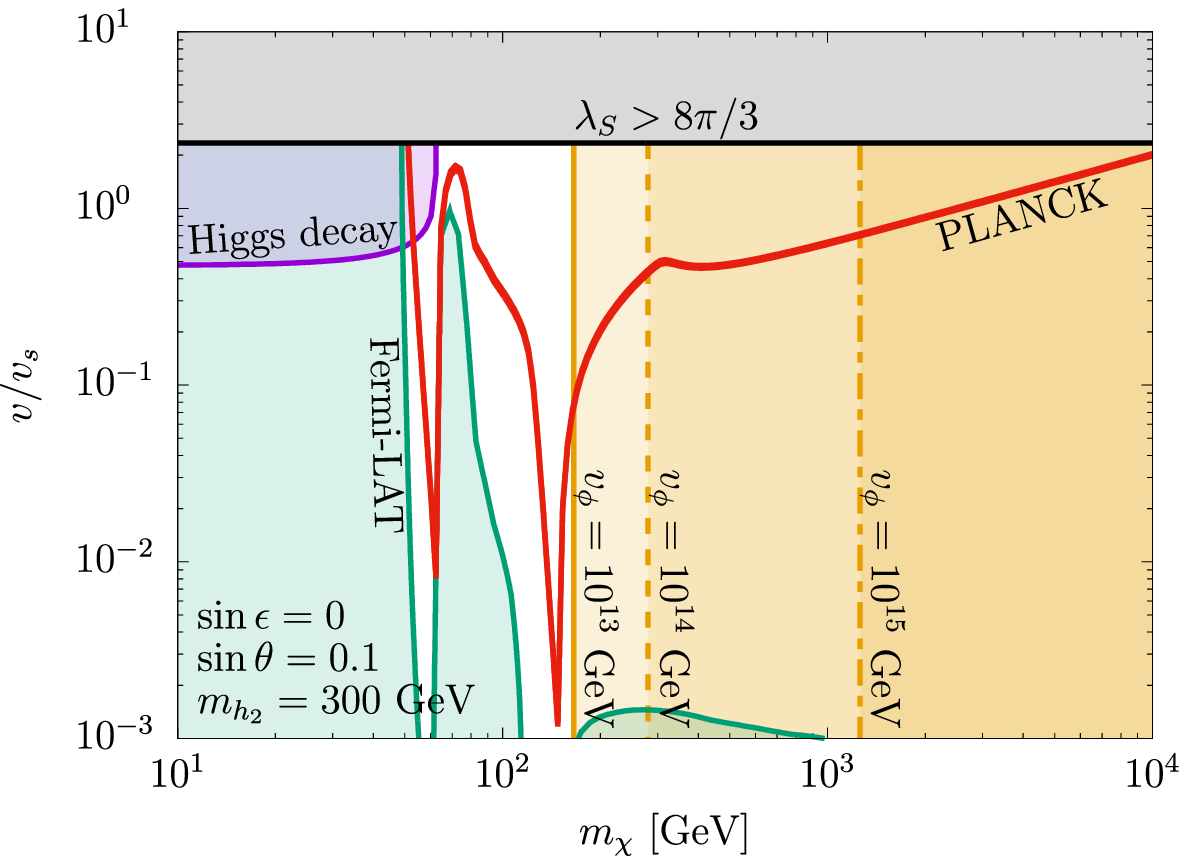}
 \includegraphics[scale=0.66]{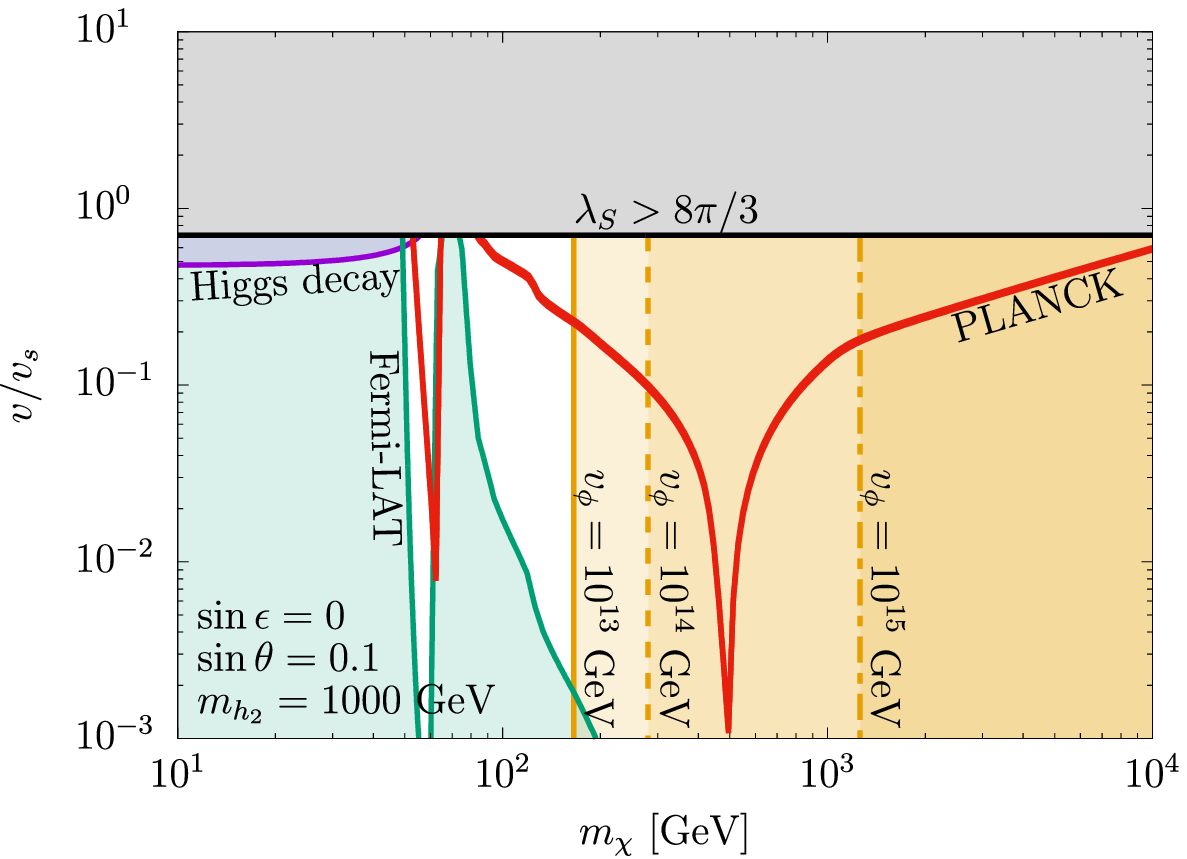}
 \\
 \includegraphics[scale=0.66]{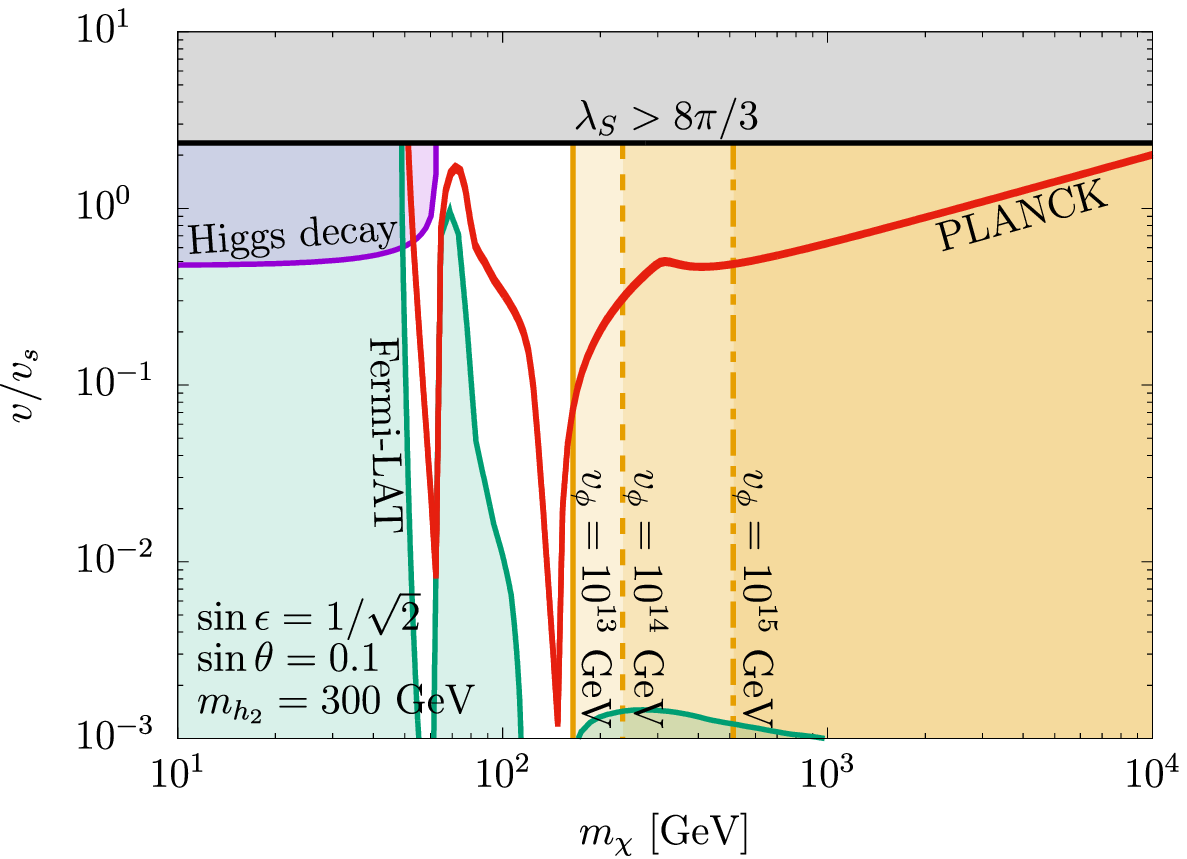}
 \includegraphics[scale=0.66]{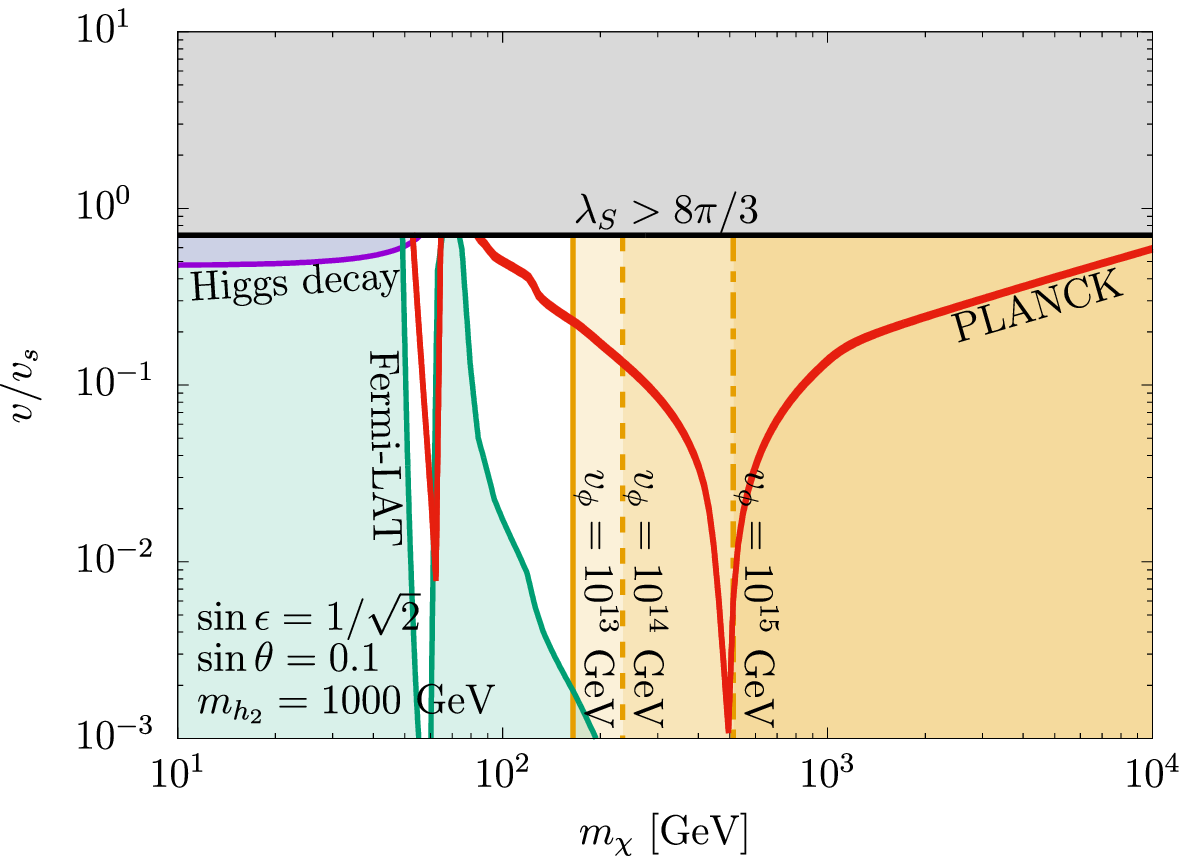}
\caption{
Allowed regions in the $(m_{\chi}, v /v_{s} )$ plane.
The scalar mass is fixed as $m_{h_2} = 300\ \mathrm{GeV}$ for the left
 panels and $m_{h_2} = 1000\ \mathrm{GeV}$ for the right panels. 
The gauge kinetic mixing is chosen as $\sin \epsilon = 0$ (no kinetic
 mixing) in the upper panels and $\sin \epsilon = 1/\sqrt{2}$ (maximal
 mixing) in the lower panels. 
The red line corresponds to the thermal dark matter relic abundance
 consistent with the PLANCK Collaboration~\cite{Aghanim:2018eyx}. 
The purple, gray, orange and green regions are excluded by the constraints of the Higgs invisible
decays~\cite{Sirunyan:2018owy, Aaboud:2019rtt} and the Higgs signal
strength~\cite{Khachatryan:2016vau}, the perturbative unitarity 
bound on $\lambda_S$~\cite{Chen:2014ask}, the cosmic-ray constraint $( \tau_\mathrm{DM} \gtrsim
10^{27}\ \mathrm{s})$~\cite{Baring:2015sza} and the gamma-ray observation~\cite{Fermi-LAT:2016uux}, respectively. 
}
\label{fig:relic}
\end{figure}

Finally, we confirm the consistency of our model with the observed dark matter 
relic abundance. 
For calculations of the dark matter relic abundance, 
the model is implemented in CalcHEP~\cite{Belyaev:2012qa} by using LanHEP~\cite{Semenov:2014rea}.
The physical quantities relevant to dark matter such as thermal relic
abundance, all the decay widths, spin-independent cross section for
direct detection are computed by using micrOMEGAs~\cite{Belanger:2018mqt}.
In Fig.~\ref{fig:relic}, we show the consistency of our pNGB dark matter model 
with the observed relic abundance in the plane $(m_{\chi}, v/v_{s})$.
The red line represents the parameter space reproducing the observed
thermal relic abundance within $3\sigma$ range of the PLANCK data
$\Omega_{\mathrm{DM}} h^2 = 0.120 \pm 0.001$~\cite{Aghanim:2018eyx}.
One can see the two resonances in Fig.~\ref{fig:relic} due to the two
Higgs bosons $h_1$ and $h_2$. 
The purple region is excluded by the measurements of the Higgs invisible
decay~\cite{Sirunyan:2018owy, Aaboud:2019rtt} and the signal
strength~\cite{Khachatryan:2016vau} and the upper gray region is ruled 
out by the perturbative unitarity bound of the quartic coupling $\lambda_S <
8\pi /3$~\cite{Chen:2014ask}.
The green region is excluded by the gamma-ray observation coming from
dwarf spheroidal galaxies where the effective annihilation cross section
into $b\overline{b}$ defined by 
$\langle\sigma_\mathrm{eff}{v}_{\mathrm{rel}}\rangle\equiv\langle\sigma_{b\bar{b}} v_{\mathrm{rel}}\rangle \left(\Omega_\mathrm{DM}h^2/0.120\right)^2$
with the dark matter relative velocity $v_{\mathrm{rel}}$
becomes larger than the current upper bound given by Fermi-LAT~\cite{Fermi-LAT:2016uux}.\footnote{
Note that the parameter space excluded by the gamma-ray observation shown in
Fig.~\ref{fig:relic} is different
from the previous work~\cite{Huitu:2018gbc}.
This is because in the previous work the dark matter
abundance in our galaxy has been assumed to be the
observed value ($\Omega_\mathrm{DM}h^2\approx0.120$) regardless of the
thermal abundance computed at each parameter space.
This can occur after thermal production of dark matter via
additional non-thermal dark matter production or entropy production, for instance. 
On the other hand in our case, thermal dark matter production is
only assumed.  
}
%
These behavior is basically same with the previous work as
expected~\cite{Gross:2017dan}.
The orange region is excluded by the upper bound on the dark matter
lifetime $\tau_\mathrm{DM}\gtrsim10^{27}~\mathrm{s}$ where the VEV is fixed as $v_{\phi} = 10^{13}\ \mathrm{GeV},
10^{14}\ \mathrm{GeV}, 10^{15}\ \mathrm{GeV}$.
One can observe from the plots that the bound becomes stronger for small
$v_\phi$ and non-zero gauge kinetic mixing $\sin\epsilon$.
%

\section{Conclusion}
We have studied the pNGB dark matter scenario derived from 
the gauged $U(1)_{B-L}$ symmetry. 
The model is consist of particles in the ordinary $U(1)_{B-L}$ model with 
an additional scalar singlet with $Q_{B-L}=+1$. 
The small neutrino masses have also been generated via type-I seesaw mechanism 
as usual. 
In this model, the pNGB associated with $U(1)_{B-L}$ symmetry breaking 
is identified as a dark matter candidate. 
The interactions of the new $U(1)_{B-L}$ gauge boson and the
scalar mixing have led the decays of the pNGB. 
We have shown that the lifetime of the pNGB is long enough to be dark matter. 
We have also found the parameter space, which are consistent 
with the relevant constraints such as observed relic abundance of dark matter,
Higgs invisible decay, Higgs signal strength, and perturbative unitarity bound 
of the couplings. 
For future prospects, the planned gamma-ray observations such as
Cherenkov Telescope Array (CTA)~\cite{Carr:2015hta} and Large High
Altitude Air Shower Observatory (LHAASO)~\cite{Bai:2019khm} can 
explore the dark matter mass over $100~\mathrm{GeV}$.
In particular, the LHAASO experiment is already being operated, and
can search the dark matter mass region between 1 TeV and 100 TeV.
The upper bound on the dark matter lifetime is expected to be updated by
one order of magnitude as discussed in Ref.~\cite{He:2019bcr}. 
These upcoming experiments will be able to explore full parameter space of our gauged $U(1)_{B-L}$ pNGB dark matter.
%

\section*{Acknowledgments}
\noindent
TT acknowledges funding from the Natural Sciences and Engineering Research Council of Canada (NSERC).
The work of KT is supported 
by the MEXT Grant-in-Aid for Scientific Research on Innovation Areas (Grant No. 18H05543). 
Numerical computation in this work was carried out at the Yukawa
Institute Computer Facility and Compute Canada (Compute Ontario).
%

\appendix
\section{Gauge kinetic mixing}
\label{sec:Gauge kinetic mixing}
When the kinetic terms of the $U(1)_{Y}$ and $U(1)_{B-L}$ gauge fields are given by
%
\begin{align}
 \mathcal{L}_{GK} = - \frac{1}{4} B_{\mu\nu} B^{\mu\nu} -\frac{1}{4} X_{\mu\nu} X^{\mu\nu} -\frac{\sin \epsilon}{2} B_{\mu\nu} X^{\mu\nu},
\end{align}
%
these can be diagonalized as
%
\begin{align}
 \mathcal{L}_{GK} = -\frac{1}{4} \hat{B}_{\mu\nu} \hat{B}^{\mu\nu} - \frac{1}{4} \hat{X}_{\mu\nu} \hat{X}^{\mu\nu},
\end{align}
%
by the linear transformation
%
\begin{align}
 \left( \begin{array}{c}
 B_{\mu\nu} \\
 X_{\mu\nu} 
 \end{array}\right)
 =
 \left( \begin{array}{cc}
 1 & -\tan \epsilon \\
 0 & 1/\cos \epsilon
 \end{array}\right)
 \left( \begin{array}{c}
 \hat{B}_{\mu\nu} \\
 \hat{X}_{\mu\nu}
 \end{array}\right)
 \equiv
 V_{GK}
 \left( \begin{array}{c}
 \hat{B}_{\mu\nu} \\
 \hat{X}_{\mu\nu}
 \end{array}\right).
\end{align}
%
On the other hand, the mass matrix of the neutral gauge bosons is given by
%
\begin{align}
 \mathcal{L}_{M}
 = \frac{1}{2}
 \left( \begin{array}{ccc}
 B_\mu & W^3_\mu & X_\mu
 \end{array}\right)
 \left( \begin{array}{ccc}
 \sin^2 \theta_W m_{\tilde{Z}}^2 & - \sin \theta_W\cos \theta_W m_{\tilde{Z}}^2 & 0 \\
 - \sin \theta_W \cos \theta_W m_{\tilde{Z}}^2 & \cos^2 \theta_W m_{\tilde{Z}}^2 & 0 \\
 0 & 0 & m_{X}^2
 \end{array}\right)
 \left( \begin{array}{c}
 B^\mu \\
 W^{3\mu} \\
 X^\mu
 \end{array}\right),
\end{align}
%
where the following parameters are defined:
%
\begin{align}
 &\sin \theta_W \equiv\frac{ g_{1} }{\sqrt{g_{1}^2+g_{2}^2}},
 \hspace{1em}
 \cos \theta_W \equiv \frac{g_{2}}{\sqrt{g_{1}^2 + g_{2}^2}},
 \\
 &m_{\tilde{Z}}^2 \equiv \frac{g_{1}^2 + g_{2}^2}{4} v^2,
 \hspace{1em}
 m_{X}^2 \equiv g_{B-L}^2 (v_{s}^2 + 4 v_{\phi}^2).
\end{align}
%
In the kinetic term diagonalized base, the mass matrix of the neutral
gauge boson $\hat{M}_G^2$ is written as
%
\begin{align}
 \hat{M}_{G}^2=
 \tilde{V}_{GK}^\mathrm{T}\left(
\begin{array}{ccc}
 \sin^2 \theta_W m_{\tilde{Z}}^2 & - \sin \theta_W \cos \theta_W m_{\tilde{Z}}^2 & 0\\
 - \sin \theta_W \cos \theta_W m_{\tilde{Z}}^2 & \cos^2 \theta_W m_{\tilde{Z}}^2 & 0\\
 0 & 0 & m_{X}^2
\end{array}
 \right)\tilde{V}_{GK},
\end{align}
where $\tilde{V}_{GK}$ is given by
\begin{align}
 \tilde{V}_{GK} =\left( \begin{array}{ccc}
 1 & 0 & -\tan \epsilon \\
 0 & 1 & 0 \\
 0 & 0 & 1/ \cos \epsilon
 \end{array}\right).
\end{align}
%
The mass matrix $\hat{M}_{G}^2$ can be diagonalized by the unitary matrix
%
\begin{align}
 U_{G} =\left( \begin{array}{ccc}
 \cos \theta_W & - \sin \theta_W & 0 \\
 \sin \theta_W & \cos \theta_W & 0\\
 0 & 0 & 1
 \end{array}\right)
 \left( \begin{array}{ccc}
 1 & 0 & 0 \\
 0 & \cos  \zeta & -\sin \zeta \\
 0 & \sin \zeta & \cos \zeta
 \end{array}\right),
\end{align}
%
where the mixing angle $\zeta$ is expressed by
%
\begin{align}
 \tan 2\zeta = \frac{ - m_{\tilde{Z}}^2 \sin \theta_W \sin 2\epsilon}{m_{X}^2 -m_{\tilde{Z}}^2 (\cos^2 \epsilon - \sin^2 \theta_W \sin^2 \epsilon)}.
\end{align}
In the limit of $m_{\tilde{Z}}^2 \ll m_{X}^2$ as in our case, we can
find that $\tan 2\zeta \approx -\frac{m_{Z}^2}{m_{Z'}^2}\sin \theta_W
\sin 2 \epsilon\ll1$. 
%
As a result, the gauge eigenstates can be written in terms of the mass
eigenstates $(A_\mu, Z_\mu, Z'_\mu)$ as 
%
 \begin{align}
  \left(
  \begin{array}{c}
   B_\mu \\
   W^3_\mu \\
   X_\mu
  \end{array}\right)
 = \tilde{V}_{GK} U_{G}
 \left( \begin{array}{c}
 A_\mu \\
 Z_\mu \\
 Z'_\mu
 \end{array}\right),
 \end{align}
%
where the gauge bosons $A_\mu$, $Z_\mu$ and $Z'_\mu$ correspond to the photon, the SM-like $Z$ boson and the new massive gauge boson.
The mass eigenvalues are given by
%
\begin{align}
 m_{Z}^2 = \frac{1}{2} \Biggl[ \overline{M}^2 - \sqrt{\overline{M}^4 - \frac{4 m_{\tilde{Z}}^2 m_{X}^2}{\cos^2 \epsilon}}\Biggr],
 \quad
 m_{Z'}^2 = \frac{1}{2} \Biggl[ \overline{M}^2 + \sqrt{\overline{M}^4 - \frac{4 m_{\tilde{Z}}^2 m_{X}^2}{\cos^2 \epsilon}}\Biggr], 
\end{align}
where $\overline{M}^2$ is defined by $\overline{M}^2\equiv m_{\tilde{Z}}^2 (1 +
\sin^2 \theta_W \tan^2 \epsilon) + m_{X}^2/\cos^2 \epsilon$. 
In the limit $\epsilon \to 0$, these mass eigenvalues are reduced to the usual expressions
%
\begin{align}
 m_{Z}^2 \to  m_{\tilde{Z}}^2 = \frac{ g_1^2 + g_2^2}{4}v^2,\label{eq:z_mass}
 \quad
 m_{Z'}^2 \to  m_{X}^2 = g_{B-L}^2 (v_s^2 + 4 v_\phi^2).
\end{align}
%
One can see that Eq.~(\ref{eq:z_mass}) corresponds to the SM $Z$ boson mass.

\medskip
Finally, we will derive the interactions of these gauge bosons, which are used to evaluate the decay widths of  the pNGB dark matter.
The interactions with the dark matter come from the covariant derivative of $S$, and its expressions are given by
%
\begin{align}
 \mathcal{L}_{Z h_i \chi} =& \sum_i  g_{B-L} \frac{\sin \zeta}{\cos \epsilon} \frac{U_{si}}{\sqrt{1 + \frac{v_s^2}{4 v_\phi^2}}} Z_\mu ( h_i \partial^\mu \chi -\chi \partial^\mu h_i),
 \\
 \mathcal{L}_{Z'h_i \chi} =& \sum_i  g_{B-L} \frac{\cos \zeta}{\cos \epsilon} \frac{U_{si}}{\sqrt{1 + \frac{v_s^2}{4 v_\phi^2}}} Z'_\mu (h_i \partial^\mu \chi -\chi \partial^\mu h_i).
\end{align}
%
%
The couplings between the heavy gauge boson $Z'$ and the (axial) vector currents of the SM fermion $f$ is defined by
%
\begin{align}
 \mathcal{L}_{Z' \bar{f}f}= -Z'_{\mu} \bar{f} \gamma^\mu \Bigl[ g^{f}_{V} +g^{f}_{A} \gamma_{5} \Bigl] f,
\end{align}
%
and the explicit expression of the coefficients are given by
\begin{align}
 g^f_V = & -\frac{g_2}{2}T_3^f \sin \zeta \cos \theta_W +g_1 (Q^f_{\mathrm{em}} -T^f_3) (\sin \zeta \sin \theta_W -\cos \zeta \tan \epsilon)
 \nonumber\\
 &+ g_{B-L} Q^f_{B-L} \frac{\cos \zeta}{\cos \epsilon},
 \\
 g^f_A =\, \,&\frac{g_2}{2} T^f_3 \sin \zeta \cos \theta_W.
\end{align}
%



\begin{thebibliography}{200}


\bibitem{Corbelli:1999af} 
  E.~Corbelli and P.~Salucci,
  Mon.\ Not.\ Roy.\ Astron.\ Soc.\  {\bf 311}, 441 (2000)
  [astro-ph/9909252].

\bibitem{Sofue:2000jx} 
  Y.~Sofue and V.~Rubin,
  Ann.\ Rev.\ Astron.\ Astrophys.\  {\bf 39}, 137 (2001)
  [astro-ph/0010594].


\bibitem{Massey:2010hh} 
  R.~Massey, T.~Kitching and J.~Richard,
  Rept.\ Prog.\ Phys.\  {\bf 73}, 086901 (2010)
  [arXiv:1001.1739 [astro-ph.CO]].


\bibitem{Aghanim:2018eyx} 
  N.~Aghanim {\it et al.} [Planck Collaboration],
  arXiv:1807.06209 [astro-ph.CO].


\bibitem{Randall:2007ph} 
  S.~W.~Randall, M.~Markevitch, D.~Clowe, A.~H.~Gonzalez and M.~Bradac,
  Astrophys.\ J.\  {\bf 679}, 1173 (2008)
  [arXiv:0704.0261 [astro-ph]].






\bibitem{Akerib:2017kat}
  D.~S.~Akerib {\it et al.} [LUX Collaboration],
  Phys.\ Rev.\ Lett.\  {\bf 118} (2017) no.25,  251302
  [arXiv:1705.03380 [astro-ph.CO]].

\bibitem{Cui:2017nnn}
  X.~Cui {\it et al.} [PandaX-II Collaboration],
  Phys.\ Rev.\ Lett.\  {\bf 119} (2017) no.18,  181302
  [arXiv:1708.06917 [astro-ph.CO]].

\bibitem{Aprile:2018dbl}
  E.~Aprile {\it et al.} [XENON Collaboration],
  Phys.\ Rev.\ Lett.\  {\bf 121} (2018) no.11,  111302
  [arXiv:1805.12562 [astro-ph.CO]].



\bibitem{Freytsis:2010ne}
  M.~Freytsis and Z.~Ligeti,
  Phys.\ Rev.\ D {\bf 83} (2011) 115009
  [arXiv:1012.5317 [hep-ph]].
	

 
\bibitem{Ipek:2014gua} 
  S.~Ipek, D.~McKeen and A.~E.~Nelson,
  Phys.\ Rev.\ D {\bf 90}, no. 5, 055021 (2014)
  [arXiv:1404.3716 [hep-ph]].

\bibitem{Arcadi:2017wqi} 
  G.~Arcadi, M.~Lindner, F.~S.~Queiroz, W.~Rodejohann and S.~Vogl,
  JCAP {\bf 1803}, no. 03, 042 (2018)
  [arXiv:1711.02110 [hep-ph]].

\bibitem{Sanderson:2018lmj} 
  N.~F.~Bell, G.~Busoni and I.~W.~Sanderson,
  JCAP {\bf 1808}, no. 08, 017 (2018)
  Erratum: [JCAP {\bf 1901}, no. 01, E01 (2019)]
  [arXiv:1803.01574 [hep-ph]].

\bibitem{Abe:2018emu} 
  T.~Abe, M.~Fujiwara and J.~Hisano,
  JHEP {\bf 1902}, 028 (2019)
  [arXiv:1810.01039 [hep-ph]].

\bibitem{Abe:2019wjw} 
  T.~Abe, M.~Fujiwara, J.~Hisano and Y.~Shoji,
  arXiv:1910.09771 [hep-ph].


	
\bibitem{Barger:2010yn} 
  V.~Barger, M.~McCaskey and G.~Shaughnessy,
  Phys.\ Rev.\ D {\bf 82}, 035019 (2010)
  [arXiv:1005.3328 [hep-ph]].

\bibitem{Gross:2017dan} 
  C.~Gross, O.~Lebedev and T.~Toma,
  Phys.\ Rev.\ Lett.\  {\bf 119}, no. 19, 191801 (2017)
  [arXiv:1708.02253 [hep-ph]].


\bibitem{Fonseca:2015gva} 
  N.~Fonseca, R.~Zukanovich Funchal, A.~Lessa and L.~Lopez-Honorez,
  JHEP {\bf 1506}, 154 (2015)
  [arXiv:1501.05957 [hep-ph]].

\bibitem{Brivio:2015kia} 
  I.~Brivio, M.~B.~Gavela, L.~Merlo, K.~Mimasu, J.~M.~No, R.~del Rey and V.~Sanz,
  JHEP {\bf 1604}, 141 (2016)
  [arXiv:1511.01099 [hep-ph]].

\bibitem{Barducci:2016fue} 
  D.~Barducci {\it et al.},
  JHEP {\bf 1701}, 078 (2017)
  [arXiv:1609.07490 [hep-ph]].

\bibitem{Balkin:2017aep}
  R.~Balkin, M.~Ruhdorfer, E.~Salvioni and A.~Weiler,
  JHEP {\bf 1711} (2017) 094
  [arXiv:1707.07685 [hep-ph]].

\bibitem{Balkin:2018tma}
  R.~Balkin, M.~Ruhdorfer, E.~Salvioni and A.~Weiler,
  JCAP {\bf 1811} (2018) 050
  [arXiv:1809.09106 [hep-ph]].

\bibitem{Ruhdorfer:2019utl} 
  M.~Ruhdorfer, E.~Salvioni and A.~Weiler,
  SciPost Phys.\  {\bf 8}, 027 (2020)
  [arXiv:1910.04170 [hep-ph]].

\bibitem{Ramos:2019qqa} 
  M.~Ramos,
  arXiv:1912.11061 [hep-ph].



\bibitem{Azevedo:2018exj} 
  D.~Azevedo, M.~Duch, B.~Grzadkowski, D.~Huang, M.~Iglicki and R.~Santos,
  JHEP {\bf 1901}, 138 (2019)
  [arXiv:1810.06105 [hep-ph]].

\bibitem{Ishiwata:2018sdi} 
  K.~Ishiwata and T.~Toma,
  JHEP {\bf 1812}, 089 (2018)
  [arXiv:1810.08139 [hep-ph]].

\bibitem{Huitu:2018gbc} 
  K.~Huitu, N.~Koivunen, O.~Lebedev, S.~Mondal and T.~Toma,
  Phys.\ Rev.\ D {\bf 100}, no. 1, 015009 (2019)
  [arXiv:1812.05952 [hep-ph]].

\bibitem{Cline:2019okt} 
  J.~M.~Cline and T.~Toma,
  Phys.\ Rev.\ D {\bf 100}, no. 3, 035023 (2019)
  [arXiv:1906.02175 [hep-ph]].



\bibitem{Arina:2019tib} 
  C.~Arina, A.~Beniwal, C.~Degrande, J.~Heisig and A.~Scaffidi,
  arXiv:1912.04008 [hep-ph].



\bibitem{Banks:1988yz}
  T.~Banks and L.~J.~Dixon,
  Nucl.\ Phys.\ B {\bf 307} (1988) 93.

\bibitem{Banks:2010zn}
  T.~Banks and N.~Seiberg,
  Phys.\ Rev.\ D {\bf 83} (2011) 084019
  [arXiv:1011.5120 [hep-th]].

	

	

	
\bibitem{Baring:2015sza} 
  M.~G.~Baring, T.~Ghosh, F.~S.~Queiroz and K.~Sinha,
  Phys.\ Rev.\ D {\bf 93}, no. 10, 103009 (2016)
  [arXiv:1510.00389 [hep-ph]].


\bibitem{Bandyopadhyay:2018cwu} 
  T.~Bandyopadhyay, G.~Bhattacharyya, D.~Das and A.~Raychaudhuri,
  Phys.\ Rev.\ D {\bf 98}, no. 3, 035027 (2018)
  [arXiv:1803.07989 [hep-ph]].









	
\bibitem{Martin-Lozano:2015dja} 
  V.~Mart\'in Lozano, J.~M.~Moreno and C.~B.~Park,
  JHEP {\bf 1508}, 004 (2015)
  [arXiv:1501.03799 [hep-ph]].

\bibitem{Falkowski:2015iwa}
  A.~Falkowski, C.~Gross and O.~Lebedev,
  JHEP {\bf 1505} (2015) 057
  [arXiv:1502.01361 [hep-ph]].



\bibitem{Charnock:2016nzm}
  T.~Charnock, A.~Avgoustidis, E.~J.~Copeland and A.~Moss,
  Phys.\ Rev.\ D \textbf{93} (2016) no.12, 123503
  [arXiv:1603.01275 [astro-ph.CO]].


\bibitem{Belyaev:2012qa} 
  A.~Belyaev, N.~D.~Christensen and A.~Pukhov,
  Comput.\ Phys.\ Commun.\  {\bf 184}, 1729 (2013)
  [arXiv:1207.6082 [hep-ph]].

\bibitem{Semenov:2014rea} 
  A.~Semenov,
  Comput.\ Phys.\ Commun.\  {\bf 201}, 167 (2016)
  [arXiv:1412.5016 [physics.comp-ph]].

\bibitem{Belanger:2018mqt} 
  G.~B\'elanger, F.~Boudjema, A.~Goudelis, A.~Pukhov and B.~Zaldivar,
  Comput.\ Phys.\ Commun.\  {\bf 231}, 173 (2018)
  [arXiv:1801.03509 [hep-ph]].


\bibitem{Sirunyan:2018owy} 
  A.~M.~Sirunyan {\it et al.} [CMS Collaboration],
  Phys.\ Lett.\ B {\bf 793}, 520 (2019)
  [arXiv:1809.05937 [hep-ex]].

\bibitem{Aaboud:2019rtt} 
  M.~Aaboud {\it et al.} [ATLAS Collaboration],
  Phys.\ Rev.\ Lett.\  {\bf 122}, no. 23, 231801 (2019)
  [arXiv:1904.05105 [hep-ex]].

\bibitem{Khachatryan:2016vau} 
  G.~Aad {\it et al.} [ATLAS and CMS Collaborations],
  JHEP {\bf 1608}, 045 (2016)
  [arXiv:1606.02266 [hep-ex]].


	

\bibitem{Chen:2014ask}
  C.~Y.~Chen, S.~Dawson and I.~M.~Lewis,
  Phys.\ Rev.\ D {\bf 91} (2015) no.3,  035015
  [arXiv:1410.5488 [hep-ph]].

\bibitem{Fermi-LAT:2016uux}
  A.~Albert \textit{et al.} [Fermi-LAT and DES],
  Astrophys.\ J.\  \textbf{834}, no.2, 110 (2017)
  [arXiv:1611.03184 [astro-ph.HE]].





	



\bibitem{Carr:2015hta} 
  J.~Carr {\it et al.} [CTA Collaboration],
  PoS ICRC {\bf 2015}, 1203 (2016)
  [arXiv:1508.06128 [astro-ph.HE]].

\bibitem{Bai:2019khm}
  X.~Bai {\it et al.},
  arXiv:1905.02773 [astro-ph.HE].

\bibitem{He:2019bcr}
  D.~Z.~He, X.~J.~Bi, S.~J.~Lin, P.~F.~Yin and X.~Zhang,
  arXiv:1910.05017 [astro-ph.HE].


\end{thebibliography}
\end{document}